\documentstyle[prb, aps, fleqn]{revtex}

\begin{document}
\title{Effect of uniaxial and biaxial crystal-field potential \\
on magnetic properties of a mixed spin-1/2 and spin-1 \\
Ising model on honeycomb lattice}
\author{Jozef Stre\v{c}ka and  Michal Ja\v{s}\v{c}ur\\
\normalsize Department of Theoretical Physics and Astrophysics,
         Institute of Physics, \\
\normalsize P. J. \v{S}af\'{a}rik University,
Park Angelinum 9, 040 01 Ko\v{s}ice, Slovak Republic\\
\normalsize E-mail: jozkos@pobox.sk, jascur@kosice.upjs.sk}
\date{Submitted:\today}
\maketitle
\begin{abstract}
Magnetic properties of a mixed spin-1/2 and spin-1
Ising model on honeycomb lattice are exactly investigated within
the framework of generalized star-triangle mapping transformation.
The particular attention is focused on the effect of uniaxial and
biaxial crystal-field anisotropies that basically influence the
magnetic behaviour of the spin-1 atoms. Our results for the basic thermodynamic
quantities, as well as the dynamical time-dependent autocorrelation function
indicate the spin tunneling between the $| +1 \rangle$ and $| - 1 \rangle$
states in the magnetically ordered phase.
\end{abstract}
{\it Keywords:} uniaxial and biaxial crystal-field anisotropy;
exact solution; star-triangle transformation \\
{\it PACS:} 75.10.Hk, 05.50.+q

\section{Introduction}

Over the last few years, many non-trivial quantum phenomena have
been discovered in the low-dimensional magnetic materials.
One of the most actively studied problems in the condensed
matter physics at present is a quantum tunneling of magnetization,
i. e. the effect, which has been recently developed in a large number
of single-molecule magnets (see Ref. [1] and references therein).
By the term single-molecule magnets, one denotes the small clusters
of magnetic metal ions that usually possess an extraordinary strong
magnetic anisotropy.
Hence, the single-molecule magnets often provide very good
examples of so-called Ising-like spin systems with a strong uniaxial
magnetic anisotropy. Of course, the Ising anisotropy by itself cannot
be a source of the quantum spin tunneling experimentally
observed in these systems. It turns out, however, that this quantum
phenomenon arises in the most cases due to the higher-order crystal-field
terms. According to a number of experimental and theoretical studies
it is now quite well established, that the observed spin tunneling
originates to a major extent from the second-order biaxial
crystal-field potential, at least in Fe$_{4}$ \cite{fe4}, Fe$_8$
\cite{fe8}, Fe$_{19}$ \cite{fe19}, or Mn$_4$ \cite{mn4} compounds.

The immense interest in the magnetic properties of small magnetic
clusters shed light on the effect of single-ion anisotropy terms
$D$ (uniaxial anisotropy) and $E$ (biaxial, also called rhombic anisotropy).
In contrast to the quite well understood role of the both single-ion
anisotropies $D$ and $E$ in the small magnetic clusters (zero-dimensional
systems), the situation is much more complicated and also obscure
in one- and two-dimensional spin systems. In fact, the ground-state properties
of a spin-$S$ Ising model with the rhombic crystal-field potential $E$,
have been only recently examined by Oitmaa and von Brasch within
an effective mapping to the transverse Ising model \cite{oitmaa}.
On the basis of this effective mapping, the $T = 0$ quantum critical
point can be exactly located for the one-dimensional model, while
for the two-dimensional models they can be obtained with a high
numerical accuracy using the linked-cluster expansion method
\cite{oitmaa,pan}.
Nevertheless, the finite temperature behaviour of these models
has not been investigated in detail beyond the standard mean-field and
effective-field theories \cite{eddeqaqi}, random phase approximation
\cite{rpa}, or linked cluster expansion \cite{wang}.
It should be stressed that the biaxial anisotropy
essentially influences the magnetic properties of a large
number of polymeric molecular-based magnetic materials, too.
From the most obvious examples one could mention: NiF$_2$ \cite{moriya},
NiNO$_3$.6H$_2$O \cite{befr}, Ni(CH$_3$COO)$_2$.4H$_2$O \cite{pofr},
Mn(CH$_3$COO)$_2$.3H$_2$O \cite{kambik}, CoF$_2$ \cite{lines},
CoCl$_2$.6H$_2$O \cite{Uryu} and a series of compounds
$\mbox{Fe} (\mbox{dc})_2 \mbox{X}$ \cite{Wickman}, where X stands
for halids and dc for the dithiocarbamate or diselenocarbamate groups,
respectively. 

Owing to this fact, in this article we will focus on the uniaxial and
biaxial crystal-field anisotropies affecting the magnetic
behaviour of the mixed spin-1/2 and spin-1
honeycomb lattice. By assuming an Ising-type exchange interaction
between the nearest-neighbouring spins, the model under
investigation can be exactly treated through the
generalized star-triangle mapping transformation. The
considered model thus provides a noble example of the statistical
system, which enables to study an interplay between quantum
effects and temperature in a spontaneously ordered
magnetic system. Moreover, the magnetic structure
of a mixed-spin honeycomb lattice occurs rather frequently also in
the molecular magnetism, what clearly demonstrates a large family of
polymeric two-dimensional compounds of chemical formula:
$\mbox{A}^{\scriptsize \mbox{I}} \mbox{M}^{\scriptsize{\mbox{II}}}
\mbox{M}^{\scriptsize{\mbox{III}}} (\mbox{C}_2 \mbox{O}_4)_3$ \cite{car},
where A$^{\scriptsize{\mbox{I}}}$ stands for a non-magnetic univalent cation
$\mbox{N}(\mbox{C}_{n} \mbox{H}_{2n+1})_4$ or
$\mbox{P}(\mbox{C}_{n} \mbox{H}_{2n+1})_4$ ($n = 3-5$),
M$^{\scriptsize{ \mbox{II}}}$ and M$^{\scriptsize{ \mbox{III}}}$ denote
two- and three-valent metal atoms
Cu$^{\scriptsize{ \mbox{II}}} (S=1/2)$, Ni$^{\scriptsize{ \mbox{II}}} (S=1)$,
Co$^{\scriptsize{\mbox{II}}} (S=3/2)$, Fe$^{\scriptsize{\mbox{II}}} (S=2)$
or Mn$^{\scriptsize{ \mbox{II}}} (S=5/2)$ and Cr$^{\scriptsize{ \mbox{III}}} (S=3/2)$
or Fe$^{\scriptsize{ \mbox{III}}} (S=5/2)$, respectively. Indeed, the crystal
structure of these polymeric molecular-based magnetic materials
consists of the well-separated two-dimensional layers in which regularly
alternating M$^{\scriptsize{ \mbox{II}}}$ and M$^{\scriptsize{ \mbox{III}}}$
magnetic metal atoms constitute more or less regular honeycomb lattice (Fig. 1).
As a consequence of the anisotropic crystalline structure of these materials,
one should also expect a relatively strong uniaxial (Ising-like) anisotropy,
as it has already been suggested in the theoretical studies
based on the effective-field theory and Monte-Carlo simulations \cite{mc}.
Hence, the magnetic compounds from the family of oxalates
$\mbox{A}^{\scriptsize \mbox{I}} \mbox{M}^{\scriptsize{\mbox{II}}}
\mbox{M}^{\scriptsize{\mbox{III}}} (\mbox{C}_2 \mbox{O}_4)_3$
represent good candidates to be described by the proposed model.

The outline of this paper is as follows. In the next section the
detailed description of the model system will be presented
and then, some basic aspects of the transformation method will be
shown. Section 3 deals with the physical interpretation of the
most interesting results and finally, some concluding remarks are
drawn in Section 4.

\section{Model and method}

Let us consider the magnetic structure of a mixed-spin
honeycomb lattice schematically depicted in Fig. 1.
To ensure exact solvability of the model under investigation,
we will further suppose that the sites of sublattice $A$ are occupied
by the spin-1/2 atoms (depicted as full circles), in contrast to the sites
of sublattice $B$ that are occupied by the spin-1 atoms (open circles).
By assuming the Ising-type exchange interaction $J$ between nearest-neighbouring
spins, the total Hamiltonian of the system takes the following form:
\begin{equation}
\hat {\cal H} = J \sum_{\langle k, j \rangle}^{3N} \hat S_k^z \hat \mu_j^z
              + D \sum_{k \in B}^{N} (\hat S_k^z)^2
              + E \sum_{k \in B}^{N} [(\hat S_k^x)^2 - (\hat S_k^y)^2],
\label{eq1}
\end{equation}
where $N$ is a total number of sites at each sublattice,
$\hat \mu_j^z$ and $\hat S_k^{\alpha} (\alpha = x, y, z)$
denote the standard spatial components of the spin-1/2 and spin-1
operators, respectively. The first summation in Eq. (\ref{eq1})
is carried out over the nearest-neighbouring spin pairs, while
the other two summations run over the sites of sublattice $B$.
Apparently, the last two terms $D$ and $E$ are the crystal-field
potentials that measure a strength of the uniaxial and biaxial
anisotropy acting on the spin-1 atoms.
It is also worth noticing that there is
one-to-one correspondence between the Hamiltonian (\ref{eq1})
and the effective spin Hamiltonian with three different
single-ion anisotropies $D^x$, $D^y$ and $D^z$:
\begin{equation}
\hat {\cal H} = J \sum_{\langle k, j \rangle}^{3N} \hat S_k^z \hat \mu_j^z
              + D^z \sum_{k \in B}^{N} (\hat S_k^z)^2
              + D^x \sum_{k \in B}^{N} (\hat S_k^x)^2
              + D^y \sum_{k \in B}^{N} (\hat S_k^y)^2.
\label{eq2}
\end{equation}
In fact, one can easily prove the equivalence between the two effective spin Hamiltonians (up to the unimportant additive constant, for a comparison see \cite{xyz}), which can be achieved using this simple mapping between
the relevant parameters included in the Hamiltonians (\ref{eq1}) and
(\ref{eq2}), respectively:
\begin{equation}
D = D^z - \frac12 (D^x + D^y), \qquad \mbox{and} \qquad E = \frac12 (D^x - D^y).
\label{eq3}
\end{equation}
It should be also mentioned here that by neglecting the biaxial
anisotropy, i. e. setting $E = 0$ in Eq. (\ref{eq1})
or equivalently $D^x = D^y$ in Eq. (\ref{eq2}),
our model reduces to the exactly soluble model of Gon\c{c}alves \cite{bc}.
Accordingly, in this work we will in particular examine the effect of
biaxial anisotropy on the thermodynamical and dynamical
properties of the model under consideration.
Nevertheless, the $E$ term emerging in the Hamiltonian (\ref{eq1}) should cause
non-trivial quantum effects, since it introduces the $x$ and $y$
components of spin operators into the Hamiltonian and thus,
it is responsible for the onset of local quantum fluctuations that
are obviously missing in the Ising model with the uniaxial crystal-field potential
$D$ only.

It is therefore of interest to discuss the origin of biaxial anisotropy.
The origin of this anisotropy term consists in the low-symmetry crystal field
of ligands from the local neighbourhood of spin-1 atoms. A threefold symmetry
axis oriented perpendicular to the honeycomb layer, however, prevents
the appearance of biaxial crystal-field potential in a regular honeycomb
lattice with a perfect arrangement of the oxalato groups, as
well as magnetic metal atoms. On the other hand, the small lattice distortion,
which occurs rather frequently in the low-dimensional polymeric compounds
due to the Jahn-Teller effect, can potentially lower the local symmetry. In
consequence of that, the distortion of lattice parameters can be regarded
as a possible source of the biaxial anisotropy. The most obvious
example, where the lattice distortion removes the threefold
symmetry axis represents the single-molecule magnet Fe$_4$,
in which three outer Fe atoms occupy two non-equivalent
positions around one central Fe atom \cite{fe4}.

Let us turn our attention to the main points of the
transformation method, which enables an exact treatment
of the model under investigation. Firstly, it is very convenient
to write the total Hamiltonian (\ref{eq1}) as a sum
of the site Hamiltonians $\hat {\cal H}_{k}$:
\begin{equation}
\hat {\cal H} = \sum_{k \in B}^{N} \hat {\cal H}_{k},
\label{eq4}
\end{equation}
where each site Hamiltonian $\hat {\cal H}_k$ involves all interaction
terms associated with the appropriate spin-1 atom residing on the $k$th
site of sublattice $B$:
\begin{equation}
\hat {\cal H}_{k} = \hat S_{k}^{z} E_k
                  + (\hat S_{k}^{z})^2 D
                  + [(\hat S_{k}^{x})^2 - (\hat S_{k}^{y})^2] E,
\label{eq5}
\end{equation}
with $E_k  =  J (\hat \mu_{k1}^{z} + \hat \mu_{k2}^{z} + \hat \mu_{k3}^{z})$.
While the Hamiltonians (\ref{eq5}) at different sites commute with each other
($[\hat {\cal H}_i, \hat {\cal H}_j] = 0$, for each $i \neq j$),
the partition function of the system can be partially factorized
and consequently, rewritten in the form:
\begin{equation}
{\cal Z} = \displaystyle{\mbox{Tr}_{\{\mu \}}}
\prod_{k = 1}^{N} \mbox{Tr}_{S_k} \exp(- \beta \hat {\cal H}_k).
\label{eq6}
\end{equation}
In above, $\beta = 1/(k_B T)$, $k_B$ is Boltzmann's constant,
$T$ the absolute temperature, $\mbox{Tr}_{ \{ \mu \} }$ means
a trace over the spin degrees of freedom of sublattice $A$ and
$\mbox{Tr}_{S_k}$ stands for a trace over the spin states
of $k$th spin from sublattice $B$. So, a crucial step in our
procedure represents the calculation of the expression
$\mbox{Tr}_{S_k} \exp(- \beta \hat {\cal H}_k)$. With regard to
this, let us write the site Hamiltonian (\ref{eq5}) in an usual
matrix representation:
\begin{eqnarray}
\hat {\cal H}_{k} =
\left(
\begin{array}{ccc}
D + E_k &  0  &   E \\
  0     &  0  &   0 \\
  E     &  0  & D - E_k  \\
\end{array}
\right),
\label{eq7}
\end{eqnarray}
in a standard basis of functions $| \pm 1 \rangle, | 0 \rangle$
corresponding, respectively, to the three possible spin states
$S_k^z = \pm 1, 0$ of $k$th atom from sublattice $B$.
Obviously, it is easy to find eigenvalues of the site Hamiltonian
(\ref{eq7}), however, with respect to further calculation,
it is more favourable to obtain directly the matrix elements of
the expression $\exp( - \beta \hat {\cal H}_k)$. Using the well-known
Cauchy integral formula, one readily obtains the matrix elements for
an arbitrary exponential function of the site Hamiltonian (\ref{eq7}):
\begin{eqnarray}
\exp(\alpha \hat {\cal H}_{k}) =
\exp(\alpha D)
\left(
\begin{array}{ccc}
\cosh(\alpha \Theta) + \frac{E_k}{\Theta} \sinh(\alpha \Theta) &  0  & \frac{E}{\Theta} \sinh(\alpha \Theta) \\
  0     &  \exp(- \alpha D)  &   0 \\
\frac{E}{\Theta} \sinh(\alpha \Theta) &  0  & \cosh(\alpha \Theta) - \frac{E_k}{\Theta} \sinh(\alpha \Theta)  \\
\end{array}
\right),
\label{eq8}
\end{eqnarray}
where $\Theta = \sqrt{E_k^2 + E^2}$ and $\alpha$ marks any
multiplicative function. After substituting $\alpha = - \beta$
into the Eq. (\ref{eq8}), the calculation of the relevant trace
$\mbox{Tr}_{S_k} \exp(- \beta \hat {\cal H}_k)$ can be
accomplished, moreover, its explicit form immediately implies
a possibility of performing a standard star-triangle mapping
transformation:
\begin{eqnarray}
\mbox{Tr}_{S_k} \exp(- \beta \hat {\cal H}_k) \! \! \! &=& \! \! \!
1 + 2 \exp(- \beta D) \cosh \Bigl( \beta
\sqrt{J^2 (\mu_{k1}^{z} + \mu_{k2}^{z} + \mu_{k3}^{z})^2 + E^2} \Bigr) =
\nonumber \\
\! \! \! &=& \! \! \!
A~\exp \Bigl[ \beta R (\mu_{k1}^{z} \mu_{k2}^{z} + \mu_{k2}^{z}
\mu_{k3}^{z} + \mu_{k3}^{z} \mu_{k1}^{z} \bigr) \Bigr],
\label{eq9}
\end{eqnarray}
which replaces the partition function of a {\it star}, i. e. the
four-spin cluster consisting of one central spin-1 atom and its three
nearest-neighbouring spin-1/2 atoms, by the partition function of
a {\it triangle}, i. e. the three-spin cluster comprising of three
spin-1/2 atoms in the corners of equilateral triangle (see Fig. 1).
The physical meaning of the mapping (\ref{eq9}) is to
remove all interaction parameters associated with the central spin-1
atom and to replace them by an effective interaction $R$ between
the outer spin-1/2 atoms. It is noteworthy, that the both mapping
parameters $A$ and $R$ are "self-consistently" given by the transformation
equation (\ref{eq9}), which must be valid for any combination of spin
states of three spin-1/2 atoms. In consequence of that one obtains:
\begin{eqnarray}
A~= \Bigl(\Phi_{1} \Phi_{2}^3 \Bigr)^{1/4}, \qquad \qquad
\beta R = \mbox{ln} \Bigl( \frac{\Phi_1}{\Phi_2} \Bigr),
\label{eq10}
\end{eqnarray}
where we have introduced the functions $\Phi_1$ and $\Phi_2$
to write the transformation parameters (\ref{eq10}) in more
abbreviated and elegant form:
\begin{eqnarray}
\Phi_{1} \! \! \! &=& \! \! \!
1 + 2 \exp(- \beta D) \cosh \Bigl( \beta
\sqrt{\bigl(3J/2)^2 + E^2} \Bigr), \nonumber \\
\Phi_{2} \! \! \! &=& \! \! \!
1 + 2 \exp(- \beta D) \cosh \Bigl( \beta
\sqrt{\bigl(J/2)^2 + E^2} \Bigr).
\label{eq11}
\end{eqnarray}

When the mapping (\ref{eq9}) is performed at each site of
the sublattice $B$, the original mixed-spin honeycomb lattice is
mapped onto the spin-1/2 triangular lattice with the effective
interaction $R$ given by the "self-consistency" condition
(\ref{eq10})-(\ref{eq11}). As a matter of fact, the substitution
of the mapping transformation (\ref{eq9}) into the partition
function (\ref{eq6}) establishes the relationship:
\begin{equation}
{\cal Z} (\beta, J, D, E) = A^{N} {\cal Z}_{t} (\beta, R),
\label{eq12}
\end{equation}
between the partition function ${\cal Z}$ of the mixed-spin honeycomb
lattice and the partition function ${\cal Z}_t$ of the corresponding
spin-1/2 triangular lattice. Above equation
constitutes the basic result of our calculation, since it enables
relatively simple derivation of all required quantities such as
magnetization, quadrupolar moment, correlation function,
internal energy, specific heat, etc. Moreover, by combining (\ref{eq12})
with (\ref{eq9}) one easily proves the validity of
following exact spin identities:
\begin{eqnarray}
\langle f_1 (\mu_i^z, \mu_j^z, ..., \mu_k^z) \rangle
\! \! \! &=& \! \! \!
\langle f_1 (\mu_i^z, \mu_j^z, ..., \mu_k^z) \rangle_t,
\nonumber \\
\langle f_2 (S_k^x, S_k^y, S^z_k, \mu_{k1}^z, \mu_{k2}^z, \mu_{k3}^z) \rangle
\! \! \! &=& \! \! \!
\Biggl \langle
\frac{\mbox{Tr}_{S_k} f_2 (S_k^x, S_k^y, S^z_k, \mu_{k1}^z, \mu_{k2}^z, \mu_{k3}^z)
\exp(- \beta \hat {\cal H}_k)}{\mbox{Tr}_{S_k} \exp(- \beta \hat {\cal H}_k)}
\Biggr \rangle,
\label{eq13}
\end{eqnarray}
where $\langle ... \rangle$ represents the standard canonical
average over the ensemble defined by the Hamiltonian (\ref{eq1}) and
$\langle ... \rangle_t$ canonical average performed on the spin-1/2
Ising triangular lattice with the effective exchange interaction $R$
(\ref{eq10})-(\ref{eq11}).
Furthermore, $f_1$ is an arbitrary function of the spin variables
belonging to the sublattice $A$, while $f_2$ denotes an arbitrary
function depending on the $k$th spin from sublattice $B$ and
its three nearest-neighbours from sublattice $A$. Applying the
first of spin identities (\ref{eq13}), one straightforwardly
attains the following results:
\begin{eqnarray}
m_A \! \! \! &\equiv& \! \! \! \langle \hat \mu_{k1}^z \rangle
    = \langle \hat \mu_{k1}^z \rangle_t \equiv m_t, \\
c_A \! \! \! &\equiv& \! \! \! \langle \hat \mu_{k1}^z \hat \mu_{k2}^z \rangle
      = \langle \hat \mu_{k1}^z \hat \mu_{k2}^z \rangle_t \equiv c_t,  \\
t_A \! \! \! &\equiv& \! \! \!
\langle \hat \mu_{k1}^z \hat \mu_{k2}^z \hat \mu_{k3}^z \rangle
  = \langle \hat \mu_{k1}^z \hat \mu_{k2}^z \hat \mu_{k3}^z  \rangle_t
\equiv t_t,
\label{eq14}
\end{eqnarray}
while the second of spin identities (\ref{eq13}) enables a derivation
of quantities depending on the spin variable from sublattice $B$, as well:
\begin{eqnarray}
m_B \! \! \! &\equiv& \! \! \! \langle \hat S_k^z \rangle =
- 3 m_A (K_1 + K_2)/2 - 2 t_A (K_1 - 3 K_2),  \\
q_B^x \! \! \! &\equiv& \! \! \! \langle (\hat S_k^x)^2 \rangle =
(K_5 + 3 K_6)/4 + 3 c_A (K_5 - K_6), \\
q_B^y \! \! \! &\equiv& \! \! \! \langle (\hat S_k^y)^2 \rangle =
(K_7 + 3 K_8)/4 + 3 c_A (K_7 - K_8),  \\
q_B^z \! \! \! &\equiv& \! \! \! \langle (\hat S_k^z)^2 \rangle =
(K_3 + 3 K_4)/4 + 3 c_A (K_3 - K_4).
\label{eq145}
\end{eqnarray}
In above, $m_A$ ($m_B$) labels the single-site magnetization at
sublattice $A$ ($B$), $q_B^{\alpha} (\alpha = x, y, z)$ are
different spatial components of quadrupolar moment and finally,
$c_A$ and $t_A$ static pair and triplet
correlation functions between the relevant spins of sublattice $A$,
respectively. Obviously, an exact solution for the both sublattice
magnetization and quadrupolar moment require the knowledge
of the single-site magnetization $m_t$, nearest-neighbour pair
correlation function $c_t$ and triplet correlation function
$t_t$ on the corresponding spin-1/2 triangular lattice unambiguously
given by (\ref{eq10})-(\ref{eq11}). Fortunately, the exact solution
for these quantities on spin-1/2 triangular lattice are known long
time ago, hence, one can utilize the final results from
references \cite{pot}. Finally, the coefficients emerging
in the previous set of Eqs. (17)-(20) are listed below:
\begin{eqnarray}
K_1 \! \! \! &=& \! \! \! F_1 (3J/2),  \qquad K_2 = F_1 (J/2), \qquad
K_3 = F_2 (3J/2), \qquad K_4 = F_2 (J/2),\nonumber \\
K_5 \! \! \! &=& \! \! \! F_3 (3J/2, -E), \quad \! K_6 = F_3 (J/2, -E), \quad \!
K_7 = F_3 (3J/2, E), \quad \! K_8 = F_3 (J/2, E),
\label{eq15}
\end{eqnarray}
where we have defined the functions $F_1 (x)$, $F_2 (x)$ and $F_3 (x, y)$
as follows:
\begin{eqnarray}
F_1 (x) \! \! \! &=& \! \! \!
\frac{x}{\sqrt{x^2 + E^2}}
\frac{2 \sinh(\beta \sqrt{x^2 + E^2})}
     {\exp(\beta D) + 2 \cosh(\beta \sqrt{x^2 + E^2})}; \nonumber \\
F_2 (x) \! \! \! &=& \! \! \!
\frac{2 \cosh(\beta \sqrt{x^2 + E^2})}
     {\exp(\beta D) + 2 \cosh(\beta \sqrt{x^2 + E^2})}; \nonumber \\
F_3 (x, y) \! \! \! &=& \! \! \!
\frac{\exp(\beta D) + \cosh(\beta \sqrt{x^2 + y^2})}
     {\exp(\beta D) + 2 \cosh(\beta \sqrt{x^2 + y^2})}
+ \frac{y}{\sqrt{x^2 + y^2}}
  \frac{\sinh(\beta \sqrt{x^2 + y^2})}
       {\exp(\beta D) + 2 \cosh(\beta \sqrt{x^2 + y^2})}.
\label{eq155}
\end{eqnarray}

At the end of this section, we will also provide an exact result for
one dynamical quantity - time-dependent autocorrelation function. It
should be noted here that the exactly soluble models offer only seldom the
possibility to investigate their spin dynamics. On the other hand, the dynamical
quantities such as autocorrelation and correlation functions are important
also from the experimental point of view, because their magnitude
directly determines the scattering cross section measured in the inelastic
neutron scattering experiments \cite{nse}, or the spin-lattice relaxation
rate provided by the nuclear magnetic resonance (NMR) techniques \cite{nmr}.
In this work, an exact treatment for the time-dependent autocorrelation
function will be elaborated. As a starting point for the calculation of
the autocorrelation function $C_{auto}^{zz} (t)$ can for
convenience serve the second of exact spin identities (\ref{eq13}):
\begin{eqnarray}
C_{auto}^{zz} (t) \! \! \! &\equiv& \! \! \!  \frac12 \langle
\hat S^z_k (0) \hat S^z_k (t) + \hat S^z_k (t) \hat S^z_k (0)
\rangle = \nonumber \\
\! \! \! &=& \! \! \! \frac12 \Bigl \langle
\frac{\mbox{Tr}_{S_k} \{ [\hat S^z_k (0) \hat S^z_k (t) +
\hat S^z_k (t) \hat S^z_k (0)] \exp(- \beta \hat {\cal H}_k) \}}
     {\mbox{Tr}_{S_k} \exp(- \beta \hat {\cal H}_k)}
\Bigr \rangle,
\label{eq16}
\end{eqnarray}
where the symmetrized form in the definition of $C_{auto}^{zz}$ is used
to construct a hermitian operator,
$\hat S^z_k (t) = \exp(\frac{i t \hat {\cal H}_k}{\hbar}) \hat S^z_k \exp(- \frac{i t \hat {\cal H}_k}{\hbar})$
represents the Heisenberg picture for the time-dependent operator $\hat S^z_k (t)$,
$\hbar$ stands for Planck's constant and $i = \sqrt{-1}$.
Next, the matrix representation of $\exp(\pm \frac{i t \hat {\cal H}_k}{\hbar})$
can be readily obtained by putting $\alpha = \pm \frac{i t}{\hbar}$ into
Eq. (\ref{eq8}). Then, after straightforward but a little bit tedious
calculation, one arrives to the final result for the dynamical
autocorrelation function:
\begin{eqnarray}
C_{auto}^{zz} (t) \! \! \! &=& \! \! \!
K_3 \Bigl( \frac14 + 3 c_t \Bigr)
\frac{(\frac{3}{2}J)^2 + E^2 \cos \Bigl(\frac{2t}{\hbar} \sqrt{(\frac{3}{2}J)^2 + E^2} \Bigr)}
     {(\frac{3}{2}J)^2 + E^2} + \nonumber \\
\! \! \! &+& \! \! \! K_4 \Bigl( \frac34 - 3 c_t \Bigr)
\frac{(\frac{1}{2}J)^2 + E^2 \cos \Bigl(\frac{2t}{\hbar} \sqrt{(\frac{1}{2}J)^2 + E^2} \Bigr)}
     {(\frac{1}{2}J)^2 + E^2}.
\label{eq17}
\end{eqnarray}

\section{Results and discussion}

Before proceeding to the discussion of the most interesting
results, it is noteworthy, that the results derived in the
previous section are rather general, i. e. they are valid for
the ferromagnetic ($J < 0$), as well as ferrimagnetic ($J > 0$)
version of the model under consideration. In what follows,
we will restrict ourselves to the analysis of the ferrimagnetic
model only, since the polymeric compounds from the family of
oxalates \cite{car} fall mostly into the class of ferrimagnets.
Nevertheless, it appears worthwhile to say that magnetic
behaviour of the ferrimagnetic system completely resembles
that one of the ferromagnetic system.
Finally, it should be emphasized that the mapping (\ref{eq9})
remains invariant under the transformation $E \leftrightarrow -E$.
As a result, one may consider without loss of generality the parameter
$E \geq 0$ and consequently, $x$-, $y$- and $z$-axis then represent
the hard-, medium- and easy-axis for a given system.

\subsection{Ground-state properties}

At first, we will take a closer look at the ground-state
behaviour. Taking into account the zero-temperature
limit $T \to 0^{+}$, one finds following condition
for a first-order phase transition line separating
the magnetically ordered phase (OP) from the disordered
phase (DP):
\begin{equation}
\frac{D}{J} = \sqrt{\Bigl( \frac32 \Bigr)^2 + \Bigl( \frac{E}{J} \Bigr)^2}.
\label{eq175}
\end{equation}
From Eqs. (\ref{eq14})-(\ref{eq155}), moreover, one easily attains
analytical results for the single-site sublattice magnetization
($m_A$, $m_B$), total single-site magnetization $m = (m_A + m_B)/2$
and different spatial components of the quadrupolar moment
$q_B^{\alpha} (\alpha = x, y, z)$ in the both phases, as well:
\begin{eqnarray}
\mbox{OP:} \qquad m_A \! \! \! &=& \! \! \! - \frac12, \qquad
m_B = \frac{\frac32}{\sqrt{\Bigl( \frac32 \Bigr)^2 + \Bigl( \frac{E}{J} \Bigr)^2}},
\qquad m = - \frac14 + \frac{\frac34}{\sqrt{\Bigl( \frac32 \Bigr)^2 + \Bigl( \frac{E}{J} \Bigr)^2}},
\nonumber \\
q_B^x \! \! \! &=& \! \! \!
\frac12 \Bigl (1 - \frac{\frac{E}{J}}{\sqrt{(\frac32)^2 + (\frac{E}{J})^2}} \Bigr), \quad
q_B^y =
\frac12 \Bigl (1 + \frac{\frac{E}{J}}{\sqrt{(\frac32)^2 + (\frac{E}{J})^2}} \Bigr), \quad
q_B^z = 1.0;
\label{eq18}
\end{eqnarray}
\begin{eqnarray}
\mbox{DP}: \qquad m_A \! \! \! &=& \! \! \! 0.0, \qquad
m_B = 0.0, \qquad m = 0.0,
\nonumber \\
q_B^x \! \! \! &=& \! \! \! 1.0, \qquad
q_B^y = 1.0, \qquad q_B^z = 0.0.
\label{eq19}
\end{eqnarray}
For better illustration, Fig. 2 depicts the ground-state
phase diagram in the $E/J$-$D/J$ plane (Fig. 2a) together with the
zero-temperature variations of the magnetization and quadrupolar
moment in the OP (Fig. 2b, the value of uniaxial anisotropy
$D/J = 0.0$ has been chosen not to pass through the phase boundary).
It is worthy to mention that by neglecting the biaxial anisotropy,
i. e. setting $E/J = 0.0$, one recovers from the phase boundary
condition (\ref{eq175}) a boundary uniaxial anisotropy $D/J = 1.5$,
which has been already reported by Gon\c{c}alves several years ago \cite{bc}.
In this limit, the OP corresponds to the simple ferrimagnetic
phase in which both sublattice magnetization are fully
saturated and also antiparallel oriented with respect to each
other (in fact, $m_A = -0.5$ and $m_B = 1.0$).

The situation becomes much more complicated by turning on the biaxial
anisotropy $E$. Even though the sublattice magnetization $m_A$
remains at its saturation value in the whole OP,
the sublattice magnetization $m_B$ is gradually suppressed by increasing
the biaxial anisotropy strength. In contrast, neither
sublattice magnetization, nor the quadrupolar moment do not depend
within either ground state phase on the uniaxial crystal-field 
potential $D$. Of course,
the relevant change of sublattice magnetization $m_B$ must reflect
a violation of a perfect ferrimagnetic spin arrangement in the OP.
To achieve the non-saturated $m_B$ at $T = 0$, some spins of sublattice $B$ must
flip from the $| +1 \rangle$ to $| -1 \rangle$ and/or $| 0 \rangle$ state(s).
It is therefore of great importance to identify the magnitude of
the quadrupolar moment $q_B^z$. Since the quadrupolar moment
approaches in the OP its saturation value $q_B^z = 1.0$ independently
of $E/J$, a presence of the $| 0 \rangle$ states can be thus clearly excluded.
These observations would suggest, that the biaxial anisotropy causes
in the OP a spin tunneling between the $| +1 \rangle$ and $| -1 \rangle$ states,
whereas the stronger the ratio $E/J$, the greater the population of the
$| -1 \rangle$ state. Anyway, the probabilities to find the spin-1 atom
in the $| \pm 1 \rangle$ state are given by these simple expressions:
$p(| \pm 1 \rangle) = \frac{1 \pm m_B}{2}$.
Altogether, the spin configuration referring to the OP at $T = 0$
can be characterized as follows:
all spin-1/2 atoms are wholly ordered in their spin {\it down} positions
($m_A = - 0.5$), while the spin-1 atoms occupy with the probability
$p(| \pm 1 \rangle)$ either the $| +1 \rangle$, or $| -1 \rangle$ state.
It should be also pointed out, that the condition $q_B^y > q_B^x$
is always satisfied when $E > 0$. This inequality between the spatial
components of quadrupolar momentum provides a confirmation, that $x$-
and $y$-axis represent under the assumption $E > 0$ the hard- and
medium-axis in the OP.

At last, let us consider the spin ordering within the DP. Interestingly,
the DP remains unaltered no matter whether the biaxial anisotropy is
zero, or not. Indeed, all spin-1 atoms occupy in the DP
exclusively the $| 0 \rangle$ state, because of $m_B = q_B^z = 0.0$.
Contrary to this, the components of quadrupolar moment perpendicular to the
$z$-axis acquire in the DP their maximum value $q_B^x = q_B^y = 1.0$.
These results can be thought as an independent check for the
scenario that accompanies the phase transition from the OP to DP: 
all spin-1 atoms indeed tending to align into the $x-y$ plane.
Accordingly, the magnetic order is completely destroyed, in fact,
the vanishing magnetization $m_A$ implies a state of complete
spin randomization at sublattice $A$. Therefore, the DP does not 
exhibit any long-range magnetic order even at $T  = 0$.

Now, another interesting question arises, namely, whether the spin-1
atoms can fluctuate in the OP between their {\it allowable} $| \pm 1 \rangle$
states. In order to obtain a reliable answer to this question, the
time-dependent autocorrelation function (\ref{eq17}) will be analysed.
In the zero-temperature limit, the dynamical autocorrelation function
$C^{zz}_{auto}$ gains after straightforward calculation:
\begin{eqnarray}
C_{auto}^{zz} (t) =
\frac{(\frac{3}{2})^2 + (\frac{E}{J})^2
\cos \Bigl(\frac{2 J t}{\hbar} \sqrt{(\frac{3}{2})^2 + (\frac{E}{J})^2} \Bigr)}
     {(\frac{3}{2})^2 + (\frac{E}{J})^2},
\label{eq20}
\end{eqnarray}
which in turn proves that $C_{auto}^{zz}$ is periodic in time with the
angular frequency $\omega_u = \frac{2 J}{\hbar} \sqrt{(\frac32)^2 + (\frac{E}{J})^2}$
and the recurrence time
$\tau = \frac{\pi \hbar}{J \sqrt{(\frac32)^2 + (\frac{E}{J})^2}}$.
According to Eq. (\ref{eq20}), the dynamical autocorrelation function
does not depend in the ground state on the uniaxial anisotropy $D$.
Owing to this fact, we will further neglect this anisotropy parameter
and set $D/J = 0.0$.
For illustrative purposes, the time variation of the autocorrelation
function $C_{auto}^{zz}$ are displayed in Fig. 3 for several values
of the biaxial anisotropy $E/J = 0.1$, $0.5$, $1.0$ and $2.0$.
It appears worthwhile to make a few remarks on foregoing results.
Since the autocorrelation function varies in time,
it clearly demonstrates the zero-temperature spin dynamics between
the {\it allowable} $| \pm 1 \rangle$ states.
From the analytical solution (\ref{eq20}) as well as
depicted behaviour one can moreover deduce a physical interpretation
of the spin dynamics, namely, the spin system necessarily recovers after
the recurrence time $\tau$ always its initial state, whereas the stronger
the ratio $E/J$, the shorter the recurrence time $\tau$.
In addition, the increasing strength of the biaxial anisotropy
enhances also the time-variation of $C_{auto}^{zz}$ (i. e. the amplitude of oscillation).
This result is taken to mean, that increasing biaxial anisotropy enlarges
also a number of the spin-1 atoms, which tunnel during the recurrence time
between the $| \pm 1 \rangle$ states. Since the equilibrium magnetization
does not varies in time, a number of atoms that tunnel from $| +1 \rangle$
to $| -1 \rangle$ state, must be definitely the same as a number of atoms
that tunnel from the $| -1 \rangle$ to $| +1 \rangle$ state.
These findings have an obvious relevance to the understanding
of the zero-temperature spin dynamics, because they enable
its explanation from the microscopic viewpoint.

\subsection{Finite-temperature behaviour}

In this part, we would like to make some comments
on the finite-temperature behaviour of the system under investigation.
Let us begin by considering the effect of uniaxial and biaxial anisotropies
on the critical behaviour. For this purpose, two typical finite-temperature
phase diagrams are illustrated in Fig. 4a and 4b. In both figures,
the OP can be located below the phase boundaries depicted as solid lines,
while above the relevant phase boundaries the usual paramagnetic phase
becomes stable. A closer mathematical analysis reveals, that the
temperature-driven phase transition between these two phases is of
second-order and belongs to the standard Ising universality class.
More specifically, Fig. 4a shows the critical temperature
as a function of the uniaxial anisotropy $D/J$ for several values
of the biaxial anisotropy $E/J$. The dependence critical temperature
versus uniaxial anisotropy is quite obvious, when increasing $D/J$,
the critical temperature tends monotonically to zero as many as the boundary
value (\ref{eq175}) is achieved. While the anisotropy term $D$ forces the
spins to lie within $x-y$ plane when $D > 0$, the $E$ term tries to align
them into $y-z$ plane. Accordingly, the increasing strength of the biaxial
anisotropy supports the magnetic long-range order related to the OP
when $D/J > 1.5$ and hence, it survives until stronger
anisotropies $D/J$. As far as the region $D < 0$ is concerned, the biaxial
anisotropy substantially lowers the critical temperature of the OP.
Apparently, this behaviour arises as a consequence of the fact, that the
$E$ term simplifies the transition between the $| \pm 1 \rangle$ states
due to the non-zero quantum fluctuations. Thus, one can conclude that
the quantum fluctuations macroscopically manifest themselves in the
reduction of the critical temperature for the easy-axis uniaxial
anisotropy (i. e. for $D < 0$, where the model Hamiltonian
(\ref{eq1}) works extremely well).

To illustrate the influence of the biaxial anisotropy on the critical
behaviour, the critical temperature versus biaxial anisotropy
dependence is shown in Fig. 4b for several values of the uniaxial
anisotropy. As one would expect, the critical temperature gradually
decreases with increasing the biaxial anisotropy strength for any $D < 0$.
In agreement with the aforementioned arguments, the appropriate
depression of the critical
temperature can be again attributed to the quantum fluctuations, which
become the stronger, the greater the ratio $E/J$. Apart from this
rather trivial finding, one also observes here the interesting dependences
with the non-monotonical behaviour of the critical temperature.
Namely, for $D/J >> 0.0$ the critical temperature firstly increases
and only then gradually decreases with the biaxial anisotropy
strength (see for instance the curve for $D/J = 1.3$).
To explain such a behaviour, it should be realized that the spin-1
atoms are preferably thermally excited to the $| 0 \rangle$ state
when $D > 0$, what means, that they are preferably excited
to the $x-y$ plane. Since the biaxial anisotropy tries
to align them into the $y-z$ plane, it favors the long-range
order along $z$-axis in that it prefers the spin
tunneling between the $| \pm 1 \rangle$ states before the population 
of the $| 0 \rangle$ one.
The most interesting result to emerge here is that there is a
strong evidence, that aforementioned argument explains an existence
of the OP even under assumption of extraordinary strong anisotropies
$D/J \geq 1.5$. In fact, the magnetic long-range order related to the
OP occurs under this condition for the strong enough biaxial anisotropies
only. Surprisingly, the magnetic long-range order results in such a peculiar 
case from the quantum fluctuations (spin tunneling) caused by the biaxial anisotropy.

Now, let us provide an independent check of the critical
behaviour by studying the thermal dependences of magnetization.
The single-site magnetization against the temperature
are plotted in Fig. 5 for the uniaxial anisotropy $D/J = -2.0$ and
several values of the biaxial anisotropy $E/J$.
Fig. 5a shows a typical situation observed by turning on the
biaxial anisotropy $E/J$: the greater this anisotropy parameter,
the stronger the reduction of sublattice magnetization $m_B$
due to the $| \pm 1 \rangle$ spin tunneling.
As it is apparent from this figure, the total magnetization exhibits
in general the standard Q-type dependences. The most striking thermal
variations of the total magnetization can be evidently found
for the biaxial anisotropies close to the value $E^0_c /J = \sqrt{27}/2$,
at which $m_A$ fully compensates $m_B$ in the ground state
(Fig. 5b, see also Fig. 2b).
It turns out, however, that all marvellous thermal dependences of the total
magnetization stemming from the identical origin - the magnetization
of sublattice $A$ is thermally more easily disturbed than the
magnetization of sublattice $B$. As a result, the P-type
dependences of total magnetization occur for $E < E^0_c$,
when the prevailing magnetization $m_B$ exhibits smaller thermal
variation than the lower magnetization $m_A$ (see cases $E/J =
2.58$ and $2.59$ in Fig. 5b). Based on our earlier remark concerning the ground-state
properties, $|m_A|$ exceeds $m_B$ if $E > E_c^0$ is satisfied.
Then, when the value of biaxial anisotropy is from the vicinity $E^0_c$,
a more rapid thermal
variation of $m_A$ results in the N-type dependence with one
compensation point. As a matter of fact, the total magnetization shows
one compensation point in which spontaneous magnetization reverses
its sign, because at lower temperatures $|m_A| > m_B$, while at higher
temperatures $|m_A| < m_B$ (see the curve for $E/J = 2.61$).
In addition, even for more stronger biaxial anisotropies
the R-type dependences of total magnetization appear ($E/J = 2.63$). 
In such a case, the total magnetization retrieves
its substandard slope from the faster thermal variation of
always dominating magnetization $m_A$.

Finally, let us proceed to the discussion of the spin dynamics
at non-zero temperatures. The time variations of the autocorrelation
function $C_{auto}^{zz}$ are plotted in Fig. 6 for three selected
values of biaxial anisotropies $E/J = 0.1$, $0.5$ and $2.0$.
To enable a comparison between the autocorrelation functions at
various $E/J$, the relevant temperatures
are normalized with respect to their critical temperatures.
It can be easily realized that the autocorrelation function is
not in general a periodic function of time at non-zero temperatures.
Indeed, $C_{auto}^{zz}$ arises according to Eq. (\ref{eq20}) as a
superposition of two harmonic oscillations - oscillation with higher
angular frequency
$\omega_u = \frac{2 J}{\hbar} \sqrt{(\frac32)^2 + (\frac{E}{J})^2}$
and another one with lower angular frequency
$\omega_l = \frac{2 J}{\hbar} \sqrt{(\frac12)^2 + (\frac{E}{J})^2}$.
The interference between these harmonic oscillations with
different frequencies and also various amplitudes
gives rise to a rather complex time variation of $C_{auto}^{zz}$,
which is in general aperiodic, displaying
nodes and other typical interference effects (see Fig. 6).
The dependences drawn in Fig. 6 nicely illustrate also the temperature effect
on the spin dynamics. Namely, it follows from these dependences,
that as the temperature increases, some amplitudes are suppressed,
while another ones become more robust. Obviously,
in the high-temperature region that amplitudes become dominant,
which coincide to the oscillation with lower angular frequency
$\omega_l$ (see lower panels in Fig. 6). In contrast, the
amplitudes arising from higher frequency oscillation $\omega_u$ dominate
at lower temperatures (see upper panels in Fig. 6).
As far as the influence of biaxial anisotropy is concerned, the stronger
the ratio $E/J$, the smaller the difference between both angular
frequencies and hence, the more expressive an interference effect between
them. It is worth mentioning that some particular biaxial anisotropies
keeping the ratio $\omega_u/\omega_l$ to be rational, what ensures
that the autocorrelation function $C_{auto}^{zz}$ is periodic in time
even at $T \neq 0$. In any other case, the $C_{auto}^{zz}$ behaves
aperiodically. As a result, in the latter case one can impose
at best some characteristic time during that the most of spins engaged
in the spin dynamics change their states, as it apparent from Fig. 6
(a1, a2, a3). Although this behaviour is quasi-periodic, its characteristic
time cannot be confused with the recurrence time $\tau$ of the former ones,
in fact, the spin system at biaxial anisotropies giving the irrational ratio
$\omega_u/\omega_l$ never approaches its initial state again.

It should be also stressed, that the uniaxial anisotropy $D$
affects the spin dynamics at $T \neq 0$, as well. To illustrate
the case, we have depicted in Fig. 7 the time variations of
autocorrelation function at $T/T_c= 1.0$ for various uniaxial
anisotropies $D/J$ and the ratio $E/J = 0.5$.
Referring to this plot, the influence of $| 0 \rangle$ states
on the spin dynamics can be understood more deeply. It is quite
evident, that a number of the $| 0 \rangle$ states becomes negligible
by taking into account the easy-axis anisotropy (e. g. $D/J = -2.0$, Fig. 7a).
Really, the quadrupolar moment $q_B^z$ becomes in this case almost
saturated, as it can be seen from the time-dependence of $C_{auto}^{zz} (t)$,
since $C_{auto}^{zz} (0) = q_B^z$.
Contrary to this, the occupation of $| 0 \rangle$ states becomes
crucial when accounting the easy-plane anisotropy
(see for instance Fig. 7d displaying the $D/J = 1.5$ case).
From the comparison of Fig. 7a-d one can conclude, that positive
(negative) anisotropy term $D$ reinforces the higher (lower) frequency
oscillation $\omega_u$ ($\omega_l$) and hence, the characteristic
time becomes considerably shorter (longer). Naturally, the
observed behaviour results from the fact, that the critical
temperature gradually falls down as the anisotropy term $D/J$ increases.
When increasing the ratio $D/J$, moreover, the oscillation amplitudes
are also suppressed, what means, that a smaller
number of spins changes during the characteristic time their spin states.
This striking feature clarifies, that the $| 0 \rangle$ states
are not engaged in the spin dynamics so greatly as the $| \pm 1 \rangle$
states, or even, they do not contribute to the spin dynamics at all.

\section{Concluding remarks}

In this article, the exact solution of the mixed spin-1/2 and spin-1
Ising model on honeycomb lattice is presented and discussed in detail.
The particular attention has been focused on the effect of
uniaxial and biaxial crystal-field potentials acting on the
spin-1 atoms. As it has been shown, the presence of the biaxial
anisotropy modifies the magnetic behaviour of studied system
in a crucial manner. It turns out that already a small amount of
the biaxial anisotropy raises a non-trivial spin dynamics and basically
influences the thermodynamic properties, as well.

The most interesting finding to emerge here constitutes an exact
evidence of the spin tunneling between the $| \pm 1 \rangle$ states 
in the magnetically ordered phase (OP). Macroscopically, the tunneling
effect decreases the critical temperature for the easy-axis uniaxial
anisotropy ($D < 0$) and also, appreciably depresses the magnetization
of spin-1 atoms from its saturation value even in the ground state
(see Eq. (\ref{eq18}) and Fig. 2b). 
The reduction of critical temperature, as well as magnetization
appears apparently due to the local quantum fluctuations arising from
the biaxial anisotropy. On the other hand, the same quantum fluctuations
can surprisingly cause an onset of the magnetic long-range order
for the extraordinary strong easy-plane anisotropies $D/J > 1.5$. 
To the best of our knowledge, such a result has not been published in the
literature before.

It should be also stressed that there is an interesting correspondence between the model described by the Hamiltonian (\ref{eq1}) and a similar Ising model with a local transverse magnetic field $\Omega$ acting on the spin-1 atoms only (for a comparison, see Ref. \cite{jascur}): 
\begin{eqnarray}
\hat {\cal H} = J \sum_{\langle k, j \rangle}^{3N} \hat S_k^z \hat \mu_j^z
              + \Omega \sum_{k \in B}^{N} \hat S_k^x.
\label{eq30}	
\end{eqnarray}
However, a similarity between the both Hamiltonians (\ref{eq1}) and (\ref{eq30}) 
is not accidental, in fact, when neglecting the uniaxial crystal-field potential $D$ in the Hamiltonian (\ref{eq1}), an effective mapping $E \leftrightarrow \Omega$ ensures the equivalence between (\ref{eq1}) and (\ref{eq30}). Since this mapping 
is not related to the magnetic structure in any fashion, the appropriate correspondence can be apparently extended to several lattice models. It is therefore valuable to mention, that the magnetic properties of lattice models 
with the local transverse field become a subject matter of many other theoretical works during the last few years \cite{tim}.
Whence, the magnetic behaviour of these systems completely resemble that
one of their counterparts with the biaxial crystal-field potential $E$ only.

Finally, let us turn back to the origin of biaxial anisotropy.  
Uprise of this anisotropy term in the mixed-spin honeycomb lattice 
is closely associated with at least a small lattice distortion.
To simplify the situation, the proposed Hamiltonian (\ref{eq1})
accounts the biaxial crystal-field anisotropy, while a difference
between exchange interactions within various spatial directions of the honeycomb lattice has been, for simplicity, omitted. Nevertheless, the developed procedure can be generalized in a rather straightforward way also to a model accounting the anisotropic interactions ($J_1, J_2, J_3$) within the three non-equivalent directions of honeycomb lattice described by the Hamiltonian:
\begin{eqnarray}
\hat {\cal H} = \sum_{\langle k, j \rangle}^{3N} J_{kj} \hat S_k^z \hat \mu_j^z
                 + D \sum_{k \in B}^{N} (\hat S_k^z)^2
                 + E \sum_{k \in B}^{N} [(\hat S_k^x)^2 - (\hat S_k^y)^2],
\label{eq31}	
\end{eqnarray}
where the nearest-neighbour exchange constant $J_{kj} = J_1$, $J_2$ or $J_3$ in dependence on which of the three non-equivalent spatial direction it deals. In addition, the biaxial anisotropy strength can be even considered as an arbitrary function (linear, quadratic, exponential, logarithmic, ...) of the ratio between appropriate interaction parameters: $E = f (J_2/J_1, J_3/J_1)$. Therefore, another interesting question arises, namely, whether the Ising model with biaxial crystal-field anisotropy taken as a function $E = f (J_2/J_1, J_3/J_1)$ can be instable with respect to the spin-Peierls phenomenon. It is quite reasonable to assume, however, that under certain conditions the energy gain from the biaxial crystal-field anisotropy exceeds the elastic energy related to the lattice deformation and hence, the biaxial anisotropy can lead to a spontaneous lattice distortion. To confirm this suggestion, our future work will be directed in this way.

{\it Acknowledgement}:
We are grateful to Oleg Derzhko and Taras Verkholyak for
stimulating discussion and useful suggestions during the
Small Triangle Meeting 2003 in Medzev. \\
This work was supported under the VEGA Grant No. 1/9034/02 and
the APVT Grant No. 20-009902.

\newpage

\begin{Large}
{\bf Figure captions}
\end{Large}
\begin{itemize}
\item [Fig. 1]
The segment of a mixed-spin honeycomb lattice. The lattice positions
of the spin-1/2 (spin-1) atoms are schematically designated by
full (open) circles, the solid lines label the interactions
between nearest-neighbouring atoms. The dashed lines represent the
effective interaction between three outer spin-1/2 atoms, which
arise after performing the mapping (\ref{eq9}) at $k$th site.
\item [Fig. 2]
a) Ground-state phase diagram in the $E/J-D/J$ plane;
b) Single-site magnetizations (full lines) and quadrupolar
moment (broken ones) versus the biaxial anisotropy $E/J$
at $T = 0$ and $D/J = 0.0$.
\item [Fig. 3]
Time variation of the dynamical autocorrelation function
$C_{auto}^{zz}$ for various values of biaxial anisotropies
$E/J = 0.1$, $0.5$, $1.0$ and $2.0$. Time axis is scaled in
$\hbar/J$ units.
\item [Fig. 4]
a) Critical temperature dependence on the uniaxial anisotropy
$D/J$ for several values of biaxial anisotropies $E/J$;
b) Critical temperature dependence on the biaxial anisotropy
$E/J$ for several values of uniaxial anisotropies $D/J$.
\item [Fig. 5]
a) Thermal dependences of the total and sublattice single-site
magnetization for $D/J = -2.0$ and $E/J = 0.0$, $1.0$ and $2.0$;
b) Various temperature dependences of the total magnetization
normalized per one site when the strength of uniaxial anisotropy
is fixed ($D/J = -2.0$) and the biaxial anisotropy varies in the
vicinity of $E_c^0$.
\item [Fig. 6]
Time variations of the dynamical autocorrelation function $C_{auto}^{zz}$
when $D/J = 0.0$ is fixed and $E/J = 0.1$, $0.5$ or $2.0$.
Upper, central and lower panels show the time variation of $C_{auto}^{zz}$
at three various temperatures, which are normalized with respect to
their critical temperatures to ensure the same ratio $T/T_c = 0.8$, $1.0$
and $1.1$, respectively.
\item [Fig. 7]
The time variations of dynamical autocorrelation function $C_{auto}^{zz}$
at critical temperature ($T/T_c = 1.0$), $E/J = 0.5$
and several values $D/J = -2.0$, $1.0$, $1.3$ and $1.5$.
\end{itemize}


\begin{thebibliography}{25}

\bibitem{QT}
R. Sessoli and D. Gatteschi, Angew. Chem. {\bf 42}, 268 (2003).

\bibitem{fe4}
A. L. Barra, A. Caneschi, A. Cornia, F. Fabrizi de Biani,
D. Gatteschi, C. Sangregorio, R. Sessoli and L. Sorace,
J. Am. Chem.  Soc. {\bf 121}, 5302 (1999);
\\
G. Amoretti, S. Carretta, R. Caciuffo, H. Casalta, A. Cornia, M.
Affronte, D. Gatteschi, Phys. Rev. B {\bf 64}, 104403 (2001).

\bibitem{fe8}
K. Wieghardt, K. Pohl, I. Jibril, G. Huttner, Angew. Chem. {\bf 96}, 63 (1984);
\\
M. Hennion, L. Pardi, I. Mirebeau, E. Suard, R. Sessoli,
A. Caneschi, Phys. Rev. B {\bf 56}, 8819 (1997);
\\
R. Sessoli, D. Gatteschi, A. Caneschi, M. A. Novak, Nature {\bf 365}, 141 (1993);
\\
R. Sessoli, Mol. Cryst. Liq. Cryst. {\bf 274}, 145 (1995).

\bibitem{fe19}
J. C. Goodwin, R. Sessoli, D. Gatteschi, W. Wernsdorfer, A. K. Powell
and S. L. Health, J. Chem. Soc. Dalton Trans., 1835 (2000);
\\
M. Affronte, J. C. Lasjaunias, W. Wernsdorfer, R. Sessoli,
D. Gatteschi, S. L. Health, A. Fort and A. Rettori, Phys. Rev. B
{\bf 66}, 064408 (2002).

\bibitem{mn4}
W. Wernsdorfer, S. Bhaduri, C. Boskovic, G. Christou and D. N.
Hendrickson, Phys. Rev. B {\bf 65}, 180403 (2002).

\bibitem{oitmaa}
J. Oitmaa and A. M. A. von Brasch, Phys. Rev. B {\bf 67}, 172402 (2003).

\bibitem{pan}
K. K. Pan and Y.-L. Wang, Phys. Rev. B {\bf 51}, 3610 (1995).

\bibitem{eddeqaqi}
N. Ch. Eddeqaqi, M. Saber, A. El-Atri and M. Kerouad,
Physica A {\bf 272}, 144 (1999);
\\
W. Jiang, G. Z. Wei and A. Du, J. Magn. Magn. Mater. {\bf 250}, 49 (2002);
\\
W. Jiang, G. Z. Wei, A. Du and L. Q. Guo, Physica A {\bf 313}, 503 (2002);
\\
W. Jiang, G. Z. Wei and Q. Zhang, Physica A {\bf 329}, 161 (2003).

\bibitem{rpa}
G. P. Taggart, R. A. Tahir-Kheli and E. Shiles, Physica {\bf 75}, 234 (1974);
\\
R. Mienas, Physica A {\bf 89}, 431 (1977).

\bibitem{wang}
K. K. Pan and Y.-L. Wang, Phys. Lett. A {\bf 178}, 325 (1993).

\bibitem{moriya}
T. Moriya, Phys. Rev. {\bf 117}, 635 (1960).

\bibitem{befr}
L. Berger and S. A. Friedberg, Phys. Rev. {\bf 136}, A158 (1964).

\bibitem{pofr}
L. G. Polgar and S. A. Friedberg, Phys. Rev. B {\bf 6}, 3497 (1972).

\bibitem{kambik}
H. Kumagai, K. \^{O}no, I. Hayashi and K. Kambe, Phys. Rev. {\bf 2}, 374 (1952).

\bibitem{lines}
M. E. Lines, Phys. Rev. {\bf 137}, A982 (1965).

\bibitem{Uryu}
N. Ury$\hat{ \mbox{u}}$, J. Skalyo and S. A. Friedberg,
Phys. Rev. {\bf 144}, 684 (1966).

\bibitem{Wickman}
G. C. DeFotis, F. Palacio and R. L. Carlin,
Phys. Rev. B {\bf 20}, 2945 (1979);
\\
G. C. DeFotis, B. K. Failon, F. V. Wells and H. H. Wickman,
Phys. Rev. B {\bf 29}, 3795 (1984).

\bibitem{car}
Z. J. Zhong, N. Matsumoto, H. \=Okawa, S. Kida, Chem. Lett., 87 (1990);
\\
H. Tamaki, M. Mitsumi, K. Nakamura, N. Matsumoto,
S. Kida, H. \=Okawa and S. Iijima, Chem. Lett., 1975 (1992);
\\
H. Tamaki, Z. J. Zhong, N. Matsumoto, S. Kida, M. Koikawa, Y. Aihiwa,
Y. Hashimoto and H. \=Okawa, J. Am. Chem. Soc. {\bf 114}, 6974 (1992);
\\
S. Iijima, T. Katsura, H. Tamaki, M. Mitsumi, N. Matsumoto
and H. \=Okawa, Mol. Cryst. Liq. Cryst. Sci. Technol. A {\bf 233}, 263 (1993);
\\
S. Decurtins, S. W. Schmalle, H. R. Ostwald, A. Linden, J.
Ensling, P. G\"{u}tlich and A. Hauser, Inorg. Chim. Acta {\bf 216}, 65 (1994);
\\
C. Mathoni\`{e}re, S. G. Carling, Y. Dou and P. Day,
J. Chem. Soc. Chem. Commun., 1554 (1994);
\\
W. M. Reiff, J. Kreisz, L. Meda and R. U. Kirss, Mol. Cryst. Liq.
Cryst. Sci. Technol. A {\bf 273}, 181 (1995);
\\
C. Mathoni\`{e}re, C. J. Nuttall, S. G. Carling and P. Day,
Inorg. Chem. {\bf 35}, 1201 (1996).

\bibitem{mc}
T. Kaneyoshi, Y. Nakamura and S. Shin,
J. Phys.: Condens. Matter {\bf 10}, 7025 (1998);
\\
Y. Nakamura, J. Phys.: Condens. Matter {\bf 12}, 4067 (2000);
\\
Y. Nakamura, Phys. Rev. B {\bf 62}, 11742 (2000);
\\
Y. Nakamura, Prog. Theor. Phys. {\bf 138}, 466 (2000);
\\
G. M. Buendia and E. Machado, J. Magn. Magn. Mater. (2004).

\bibitem{xyz}
J. Stre\v{c}ka, M. Ja\v{s}\v{c}ur,
Acta Electrotechnica et Informatica {\bf 2}, 102 (2002) (cond-mat/0207519).

\bibitem{bc}
L. L. Gon\c{c}alves, Physica Scripta {\bf 32}, 248 (1985);
\\
L. L. Gon\c{c}alves, Physica Scripta {\bf 33}, 192 (1986);
\\
J. W. Tucker, J. Magn. Magn. Mater. {\bf 95}, 133 (1999).

\bibitem{pot}
R. B. Potts, Phys. Rev. {\bf 88}, 352 (1952);
\\
G. F. Newell,  Phys. Rev. {\bf 79}, 876 (1950);
\\
R. M. F. Houtappel, Physica {\bf 16}, 425 (1950);
\\
H. N. V. Temperley, Proc. Roy. Soc. {\bf 203A}, 202 (1950);
\\
R. J. Baxter and T. Choy, Proc. Roy. Soc. London Ser. A {\bf 423}, 279 (1989);
\\
J. H. Barry, T. Tanaka, M. Khatun and C. H. M\'unera,
Phys. Rev. B {\bf 44}, 2595 (1991).

\bibitem{nse}
E. Balcar, S. W. Lovesey,
{\it Theory of magnetic neutron and photon scattering},
(Clarendon, Oxford, 1989).

\bibitem{nmr}
T. Moriya, Prog. Theor. Phys. {\bf 16}, 23 (1956).

\bibitem{jascur}
M. Ja\v{s}\v{c}ur, S. Lackov\'a,
J. Phys.: Condens. Matter {\bf 12}, L583 (2000).

\bibitem{tim}
M. Ja\v{s}\v{c}ur, J. Stre\v{c}ka, Phys. Lett. A {\bf 258}, 47 (1999); \\
J. Stre\v{c}ka, H. \v{C}en\v{c}arikov\'a, M. Ja\v{s}\v{c}ur, \\
Acta Electrotechnica et Informatica {\bf 2}, 107 (2002) (cond-mat/0207518); \\
H. Suzuki, M. Suzuki, Int. J. Mod. Phys. B {\bf 16}, 3871 (2002); \\
J. Stre\v{c}ka, M. Ja\v{s}\v{c}ur, J. Magn. Magn. Mater. {\bf 260}, 415 (2003); \\
Y. Fukumoto and A. Oguchi, J. Magn. Magn. Mater. (2004).

\end{thebibliography}
\end{document}